\documentclass[12pt]{article} 

 \def \G{G$\hspace{.01 em}\ddot{\hspace{-.01 em}\mathrm o}$del }

 \def \DSS{d\mathbf{s}^2}
 \def \GS{~$\!\mathcal{GS}$~}

 \newcommand*{\EQ}[1]{\begin{eqnarray}\label{#1}}
 \newcommand*{\EEQ}[0]{\end{eqnarray}}
 \def \NN{\nonumber \\}
 \def \Ref#1{Eq.(\ref{#1})}
 \def \T{\textstyle}
 \def \D{\displaystyle}
 \def \S{\scriptstyle}
 \def \half{\T{\frac{1}{2}}\D}
 \def \fourth{\T{\frac{1}{4}}\D}
 \def \gam{\cosh r}
 \def \sig{\sinh r}

\begin{document}
 \title{\G-type space-time metrics }
 \author{Antonio Enea Romano, Charles Goebel \\
 {\normalsize Physics Department, University of Wisconsin, } \\
 {\normalsize 1150 University Avenue, 53706 Madison, USA}}
 \maketitle

 \begin{abstract}
 A simple group theoretic derivation is given of the family of space-time
 metrics with isometry group ~$SO(2,1)\times SO(2)\times\Re$~ first
 described by \G, of which the \G stationary cosmological solution is the
 member with a perfect-fluid stress-energy tensor.  Other members of the
 family are shown to be interpretable as cosmological solutions with a
 electrically charged perfect fluid and a magnetic field.

 \end{abstract}



 \section{Introduction}

 The \G metric \cite{God1
 } was the first known cosmological solution of the Einstein field
 equations with rotating matter and closed timelike curves. By `\G-type'
 metrics we mean a one-parameter family of space-time metrics of the form
 \EQ{e1}
 ds^{2} = d\mathbf{s}^2-dz^2
 \EEQ
 where $\DSS$ is the metric of a signature (+\,--\,--) space \GS with
 isometry group ~$SO(2,1)\times SO(2)$\,.  Since $\DSS$ is independent of
 $z$  the isometry group of $ds^2$ is then ~$SO(2,1)\times SO(2)\times\Re$~
  where the factor $\Re$ is the one-dimensional translation group
 ~$z\rightarrow z\,+\,$constant. (The simply transitive subgroup ~$SO(2,1)
 \times\Re$~ of ~$SO(2,1)\times SO(2)\times\Re$~ shows that \G-type
 solutions are four-dimensionally uniform, i.e., spatially uniform and
 stationary.)

  \G \cite{God1} (footnote 12) derived \GS by starting with
 three-dimensional anti-deSitter (adS) space, a space of constant positive
 curvature and signature (+\,--\,--), realizable as a pseudosphere in a
 flat $2+2$ space, and `stretched' it by a factor $\mu$ in the direction
 of a system of time-like Clifford parallels (= Hopf lines).  [ $\mu$ is
 the `one parameter' of the family of \G-type metrics; throughout this
 paper we ignore any trivial overall (scale) factor of $ds^2$.]  The
 `stretch' reduces the the adS-space isometry ~$SO(2,1)\times SO(2,1)$~ to
 ~$SO(2,1)\times SO(2)$\,.  In section 2 we exploit the fact that the
 three-dimensional adS space is a group space, namely of $SO(2,1)$\,, to
 give an alternative derivation of $\DSS$ starting with a group-space
 metric.

 The \G cosmological solution is the one member (namely with stretch $\mu
 = \sqrt{2}\,$) of the \G-type metrics, \Ref{e1}, for which the stress
 tensor $T_{\mu\nu}$ is that of a perfect fluid.  In section 3 we show
 that other amounts of stretch yield cosmological solutions which contain
 also a uniform magnetic field and a uniform electrical charge density.

 \section{Construction of the metric tensor}

 Given a matrix representation M of a continuous group G, a metric $dl^2$
 which is left and right invariant under the action of G is given by
 \EQ{e2}
 {dl^2} = \mathrm{tr}\,(dM\,M^{-1}\,dM\,M^{-1}).
 \EEQ
 [ Of course this is the Cartan-Killing metric (up to a constant factor),
 the unique invariant quadratic form of a Lie-group space. ]
 Insertion of a matrix $H$ between $ dM $ and $M^{-1}$ gives a modified
 metric
 \EQ{e3}
 dL^2 = \mathrm{tr}\,(dM\,H\,M^{-1}dM\,H\,M^{-1})
 \EEQ
 which is still left invariant, but generally no longer right invariant.
 In fact under right multiplication ~$M\rightarrow Mg\,,~g\in G\,,$
 ~$dL^2$\, transforms to
 $$
 \mathrm
 {tr}\,(dM\,g\,H\,g^{-1}M^{-1}dM\,g\,H\,g^{-1}M^{-1})\,,
 $$
 so the only remaining right invariance of the modified metric is under
 the subgroup of $G$ which commutes with $H$.


 If we take $G$ to be $SO(2,1)$ (this is \G's `hyperbolic quaternion'
 group), then $dl^2$, \Ref{e2}, is the metric of a  2\,+\,1 adS space.  To
 obtain a metric with the isometry group ~$SO(2,1)\times SO(2)$~ it
 suffices to choose an $H$ which commutes only with an $SO(2)$ subgroup of
 $G$\,.

 Because $SO(2,1)$ is isomorphic to $SL(2,\Re)$, we can  take
 \EQ{e4}
  M=e^{\,\T\varepsilon\,\phi\S/2}\;e^{\,\T\sigma_3
 \,r\S/2}\;e^{\,\T\varepsilon\,\psi\S/2}
 \qquad,\qquad \varepsilon\equiv i\sigma_{2}
 \EEQ
 where $\sigma_{j}, ~j=1,2,3$ ~are the Pauli matrices and ~$\phi,\psi,r$~
 are real parameters.

 The general form of a matrix $H$ which commutes with the
 $SO(2)$ subgroup generated by $\varepsilon$\, is
 \EQ{e5}
 H \quad = \quad a+b\,\varepsilon \quad = \quad \left [\begin {array}{cc}
 a&b\\-b&a\end {array} \right ] \,,
 \EEQ
 with $a,b$ complex numbers.  Writing for short
 \begin{eqnarray}\label{e6}
 \gamma\equiv \gam\,,~~\sigma\equiv \sig\,, ~~~c\equiv
 \cos\phi\,,~~s\equiv \sin\phi \\
 \nonumber (\,\mathrm{note}~~~ \gamma^2-\sigma^2 = 1\,,~~c^2+s^2 =
 1\,)~~~~~~~~\nonumber
 \end{eqnarray}
 we have
 \begin{eqnarray}\label{e7}
 e^{\,\T\sigma_3 \,r\S/2}~\varepsilon~ e^{-\T\sigma_3 \,r\S/2} &=& \gamma\,\varepsilon \,+ \sigma\,\sigma_1 \NN
 e^{\,\T\varepsilon\,\phi\S/2}~\,\sigma_3\,\,e^{-\T\varepsilon\,\phi\S/2} &=& c\,\sigma_3 - s\,\sigma_1 \\
 e^{\,\T\varepsilon\,\phi\S/2}~\,\sigma_1\,\,e^ {-\T\varepsilon\,\phi\S/2}  &=& c\,\sigma_1 + s\,\sigma_3 \nonumber
 \end{eqnarray}
 and hence an easy calculation gives
 {\renewcommand{\arraystretch}{2}
 \EQ{e8} \begin{array}{cclcl}
 \D 2\frac{dM}{dr}HM^{-1} &=& a\,(c\,\sigma_3 - s\,\sigma_1) & + &
 b\,[\gamma\,(c\,\sigma_1 + s\,\sigma_3)+\sigma\,\varepsilon] \\
 \D 2\frac{dM}{d\phi}HM^{-1} &=& a\,\varepsilon & - & b\,[\gamma - \sigma\
 ,(c\,\sigma_3 - s\,\sigma_1)] \\
 \D 2\frac{dM}{d\psi}HM^{-1} &=&
 a\,[\gamma\,\varepsilon+\sigma\,(c\,\sigma_1 + s\,\sigma_3)] & - & b
 \,. \end{array} \EEQ }
 Substituting into \Ref{e3} and using
 ~tr$(\sigma_i\sigma_j) = 2\delta_{ij}$
 ~we obtain for $-4\,dL^2$
 \begin{eqnarray}
 &&(a^2-b^2)\,d\psi^2 + 2(a^2-b^2)\,\gamma\,d\psi\,d\phi + (a^2 + b^2 - 2b^
 2\gamma^2)\,d\phi^2 - (a^2+b^2)\,dr^2 \NN
 &=& (a^2-b^2)\,(d\psi + \gamma\,d\phi)^2 - (a^2+b^2)\,(\gamma^2 -
 1)\,d\phi^2 - (a^2+b^2)\,dr^2 \NN
 &\doteq& \mu^2\,(d\psi + \gamma\,d\phi)^2 - \sigma^2\,d\phi^2 - dr^2
 \qquad \mathrm{where} \qquad \mu^2 \equiv \frac{a^2-b^2}{a^2+b^2} \NN
 &=& \mu^2\,[\,d\psi + d\phi + (\gamma - 1)\,d\phi\,]^2 - \sigma^2\,d\phi^
 2 - dr^2 \NN  \label{e9}
 &=& [\,dt + \mu\,(\gamma - 1)\,d\phi\,]^2 - \sigma^2\,d\phi^2 - dr^2
 \qquad \mathrm{where} \quad dt\equiv \mu(d\psi + d\phi\,)\,.
 \end{eqnarray}
 This metric and its curvature are nonsingular at $r=0$ if $r$ and $\phi$
 are regarded as polar coordinates, i.e., $\phi$ and $\phi+2\pi$ specify
 the same point and $r$ is nonnegative.  It is the metric of the space \GS
 which \G \cite{God1} (footnote 12) obtained by stretching by the factor
 $\mu$ a three-dimensional anti-deSitter space in the direction of a
 system of time-like Clifford parallels.  [\,For ~$\mu=1$\,, hence $b=0$
 in \Ref{e5}, \Ref{e9} is the metric of the adS space; upon multiplying
 the first term by $\mu^2$ and then redefining $dt\rightarrow dt/\mu$ the
 metric \Ref{e9} for general $\mu$ is obtained.\,]
 The metric \Ref{e9} was also obtained in \cite{kill} (in their Eq.(38)
 replace $\tau$ by $t/2$ and $r$ by $r/2$\,) by what they call a `squash',
 rather than a stretch.  And finally, it is a special case of the metric
 (III.14) of \cite{tipl}, for $e=-1\,,~F~\mathrm{and}~G = \mathrm{constant}
 $\,.

 Using the metric \Ref{e9} for the $\DSS$ of \Ref{e1},
 \EQ{e10}
 ds^2 = [\,dt + \mu\,(\gamma - 1)\,d\phi\,]^2 - \sigma^2\,d\phi^2 - dr^2 -
 dz^2
 \EEQ
 is a one-parameter family of metrics with the isometry $SO(2,1)\times SO(
 2)\times\Re$\,.  The corresponding metric tensor is
 \EQ{e11}
 g_{\mu\nu}=
 \left ( \begin{array}{cccc}
 1 & \mu(\gamma - 1) & 0 & 0 \\
 \mu(\gamma - 1)
  & \mu^2(\gamma - 1)^2 - (\gamma^2 - 1)
  & 0 & 0 \\
 0 & 0 & -1 & 0 \\ 0 & 0 & 0 & -1
 \end{array} \right )
 \left | \begin{array}{c}
 {\S0}~~~t \\
 {\S1}~~~\phi \\
 {\S2}~~~r \\ {\S3}~~~z
 \end{array} \right.
 \EEQ
 The Ricci tensor of this metric is found to be
 \EQ{e12}
 R_{\mu\nu}=
 \half \left( \begin{array}{cccc}
 \mu^2 & \mu^3(\gamma - 1) & 0 & 0 \\
 \mu^3(\gamma - 1)
  & (\mu^2-2)(\gamma^2-1) + \mu^4(\gamma-1)^2
  & 0 & 0 \\
 0 & 0 &\mu^2 - 2 & 0 \\ 0 & 0 & 0 & 0
 \end{array} \right) .
 \EEQ
 This can be written as
 \EQ{e13}
 R_{mn} = (\mu^2 - 1) u_m u_n - \half (\mu^2-2)g_{mn}\quad;\quad
 R_{3\nu} = 0
 \EEQ
 where ~$R_{mn}$~ is ~$R_{\mu\nu}$ with $\mu,\nu$ restricted to the range
 $0\dots2$  ~(i.e., the Roman indices ~$m,n$~ take on the values $0,1,2$)~
 and $u^\mu$ is the unit tangent of fixed-($\phi,r,z$) lines,
 \EQ{e14}
 u^\mu = (~1~~0~~0~~0~)\quad,\quad
 u_\mu = (~1~~\mu(\gamma - 1)~~0~~0~) \,.
 \EEQ
 Note that the curl of $u_\mu$ has the nonvanishing component
 \EQ{e15}
 u_{[1,2]} = u_{1,2} = \mu\,(\gamma - 1)_{,r} = \mu\sigma\,,~~\approx\mu r
 ~~~\mathrm{for~small}~r\,,
 \EEQ
 which means that points at fixed $(\phi,r,z)$ are rotating at the angular
 velocity ~$\omega = -\half\mu$\,, where positive $\omega$ means rotation
 in the $+\phi$ direction.

 It also may be mentioned that when $\mu^2 > 1$, the metric allows closed
 timelike curves, for example `$\phi$-curves' (fixed $t,r,z$), with
 $r>\ln{\sqrt{(\mu+1)/(\mu-1)}}$\,.

 From \Ref{e13} it follows that the Ricci scalar is
 \EQ{e17}
 R = R^\nu_\nu = (\mu^2 - 1) - {\T\frac{3}{2}}(\mu^2 - 2)
  = -\half(\mu^2-4)
 \EEQ
 and the Einstein tensor  ~~$G_{\mu\nu} = R_{\mu\nu} - \half R\,g_{\mu\nu}
 $~~ is
 \EQ{e18}
 G_{mn} = (\mu^2 - 1) u_m u_n - \fourth\mu^2 g_{mn}\quad;\quad G_{3n} = 0\quad;\quad
 G_{33} = \fourth(\mu^2 - 4)g_{33}~.
 \EEQ
 If ~$\mu^2 = 2\,,~R_{\mu\nu}$\, and \,$G_{\mu\nu}$\, are of the form
 ~$C_1 u_\mu u_\nu + C_2\,g_{\mu\nu}$~ and so (as observed by \G) the
 metric is a solution of the Einstein equation ~~$G_{\mu\nu} =
 T_{\mu\nu}$~ where $T_{\mu\nu}$ is the stress tensor ~~$T_{\mu\nu} =
 (\rho + p)u_\mu u_\nu - p\,g_{\mu\nu}\,$~ of a perfect fluid where its
 four-velocity $u^\mu$ is given in \Ref{e14} and its (proper) energy
 density $\rho$ and pressure $p$ are
 \EQ{e19}
 \rho = p = \half\,.
 \EEQ
 If the perfect fluid is vacuum (`cosmological constant') plus matter then
 \\
  ~$\rho = \rho_{vac} + \rho_{matter}$~~ and ~~$p = -\rho_{vac} +
 p_{matter}$~~ so
 \EQ{e20}
 \rho_{matter} = \frac{1}{1 + f}\quad;\quad \rho_{vac} = -\half\,\frac{1 -
 f}{1 + f} \qquad \mathrm{where}\quad f = \frac{p_{matter}}{\rho_{matter}}
 \,.
 \EEQ
  If the matter is dust, f = 0, then ~~$\rho_{matter} = 1 ~;~ \rho_{vac} =
 -\half$\,.

 \section{Solutions with $\mu \not= 0$}

 We now show that when $\mu^2 > 2$, the $T^{\mu\nu}$ implied by the metric
 \Ref{e11} and its Einstein tensor \Ref{e18} can be realized by adding to
 a perfect fluid a uniform magnetic field and a uniform electric charge
 density.

 As will be evident later,
 \EQ{e21}
 F_{\mu\nu} = \left( \delta^1_\mu\,\delta^2_\nu -
 \delta^2_\mu\,\delta^1_\nu\,\right) \sigma\,B
 \qquad;\qquad B = \mathrm{constant}
 \EEQ
 describes a uniform magnetic field in the $3$ (`$z$') direction.  The
 resulting electromagnetic $T^{\mu\nu}$ is
 \begin{eqnarray}\label{e22}
 \stackrel{\mathrm{EM}}{T_{\mu\nu}} &\equiv&
 F_{\mu\rho}g^{\rho\sigma}F_{\sigma\nu} -
 \fourth g_{\mu\nu}F_{\rho\alpha}g^{\alpha\beta}
 F_{\beta\sigma}g^{\rho\sigma}\,, \NN
 &=& \left( \sigma^2\,\delta^1_\mu\,\delta^1_\nu + \delta^2_\mu\,\delta^2_\nu + \half g_{\mu\nu}\,\right) B^2\,.
 \end{eqnarray}
 By using
 \EQ{e23}
 g_{mn} = u_m\,u_n - \left(\sigma^2\,\delta^1_m\,\delta^1_n +
 \delta^2_m\,\delta^2_n\,\right) \qquad;\qquad g_{3n}=0\,,
 \EEQ
 which can be seen from \Ref{e11} and \Ref{e14}, we can write
 $\stackrel{\mathrm{EM}}{T_{\mu\nu}}$ as
 \EQ{e24}
 \stackrel{\mathrm{EM}}{T_{mn}} = (\,u_m u_n -
 \half g_{mn}\,)\,B^2 \qquad;\qquad
 \stackrel{\mathrm{EM}}{T_{3n}} = 0 \qquad;\qquad
 \stackrel{\mathrm{EM}}{T_{33}} = \half g_{33}\,B^2\,.
 \EEQ
 The total $T^{\mu\nu}$ of a perfect fluid plus the magnetic field is then
 \EQ{e25}
 T_{mn} = (\,\rho + p + B^2\,)\,u_m u_n -
 (\,p + \half B^2\,)\,g_{mn} \NN
 T_{3n} = 0 \qquad;\qquad T_{33} = -(\,p - \half B^2\,)\,g_{33}
 \EEQ
 and Einstein's equation for ~$G_{\mu\nu}$\,, \Ref{e18}, is satisfied by
 \EQ{e26}
 \rho = \half(\mu^2 - 1) \qquad;\qquad p = \half \qquad;\qquad B^2 = \half(
 \mu^2 - 2) \,.
 \EEQ
 Note that $B$ is real only if ~$\mu^2 \ge 2\,.$  Also note that if ~$\mu^
 2 \ge 4$~ then ~$\rho \ge 3p$~ and a vacuum contribution to $T^{\mu\nu}$
 is not needed.

 We now consider Maxwell's equations, which in terms of ordinary gradients
 are
 \EQ{e27}
 F_{\mu\nu,\rho} + F_{\nu\rho,\mu} + F_{\rho\mu,\nu} = 0 \qquad;\qquad
 \mathcal{F}^{\mu\nu}{}_{,\nu} = -\mathcal{J}^\mu
 \EEQ
 where ~$\mathcal{F}^{\mu\nu} = \sqrt{-g}\,F^{\mu\nu}$~ and
 ~$\mathcal{J}^\mu = \sqrt{-g}\,J^\mu$\,, ~where ~$g  \equiv\mathrm{det}\,g_{\alpha\beta}\, ~= -\sigma^2$~ for the metric \Ref{e11}.
 The first (`curl') equation, which states the vanishing of magnetic
 four-current, is obviously satisfied by the field \Ref{e21}, since the
 only nonvanishing gradient is $\partial_2~(=\partial_r\,)$~ and all
 nonvanishing $F_{\mu\nu}$ have an index `2'.  As for the `div' equation,
 since
 \EQ{e28}
 \mathcal{F}^{21} &=& \sqrt{-g}\,g^{22}\,g^{11}\,F_{21} = -B \NN
 \mathcal{F}^{20} &=& \sqrt{-g}\,g^{22}\,g^{01}\,F_{21} = \mu(\gamma - 1)B\
 ,,
 \EEQ
 the only nonvanishing component of $\mathcal{J}^\mu$ is
 \EQ{e29}
 \mathcal{J}^0 = \mathcal{F}^{20}{}_{,2} = \mu\,\sigma B\,,\qquad\mathrm{i.
 e.,}\qquad J^0 = \mu B\,,
 \EEQ
 a constant electric charge density.  The physical reason for this charge
 density is that although in the rest frame of the matter the field is
 pure magnetic, in an inertial (nonrotating) frame with spatial origin
 fixed at, say, $r=0$, there is an electric field in the radial ($dr$)
 direction, with magnitude $\sim r$ at small $r$\,, hence with
 $\nabla\cdot\mathbf{E}$ constant.  [\,It might be remarked that in a
 Friedman (homogeneous, isotropic) cosmological solution a nonvanishing
 charge density is impossible because it would imply an electric field
 with nonuniform magnitude and hence a nonuniform (inhomogeneous)
 electromagnetic stress tensor.\,]

This work was partially supported by the U.S. National Science
Foundation Grant No. PHY-0070161 at the University of Wisconsin.


 \end{document}